\documentclass[prd,preprintnumbers,nofootinbib]{revtex4} 
\usepackage{graphicx} 
\usepackage{amsmath}
\usepackage{amsfonts,amsbsy}
\usepackage{amssymb}

\usepackage{pdfpages}

\def\be{\begin{equation}}
\def\ee{\end{equation}}
\def\bea{\begin{eqnarray}}
\def\eea{\end{eqnarray}}

\def\eq#1{{Eq.~(\ref{#1})}}
\def\fig#1{{Fig.~\ref{#1}}}

\def\beq{\begin{equation}}
\def\eeq{\end{equation}}
\def\bea{\begin{eqnarray}}
\def\eea{\end{eqnarray}}
\def\eq#1{{Eq.~(\ref{#1})}}
\def\fig#1{{Fig.~\ref{#1}}}

\newcommand{\Lb}{\left(}
\newcommand{\Rb}{\right)}

\newcommand{\rt}{\boldsymbol{r}_T}
\newcommand{\bt}{\boldsymbol{b}_T}

\newcommand{\qt}{{\boldsymbol{q}_T}}
\newcommand{\kt}{\boldsymbol{k}_T}

\newcommand{\ud}{\mathrm{d}}

\begin{document}

\title{\bf Comparison of the Color Glass Condensate to multiplicity dependence of the average transverse momentum in p+p, p+Pb and Pb+Pb collisions at the LHC }

%\preprint{}

\author{Amir H. Rezaeian}
\affiliation{
 Departamento de F\'\i sica,
Centro de Estudios Subat$\acute{o}$micos,
Universidad T$\acute{e}$cnica Federico Santa Mar\'\i a,\\ and
Centro Cientifico-Tecnol$\acute{o}$gico de Valpara\'\i so,
Casilla 110-V,  Valparaiso, Chile}

\begin{abstract}
The first moment $\langle p_T\rangle$ of the charged-particle transverse momentum spectrum  and its correlation with the charged-particle multiplicity $N_{ch}$ provide vital information about the underlying particle production mechanism. The ALICE collaboration recently reported that  $\langle p_T\rangle$ versus $N_{ch}$ in Pb+Pb collisions 
is smaller than in p+p and p+Pb collisions. Other interesting features of data is rather flatness of $\langle p_T\rangle$ at high $N_{ch}$ in Pb+Pb and p+Pb collisions in seemingly striking contrast to the case of p+p collisions. With a detailed calculation, we show all these peculiar features in a wide range of energies and system sizes can be well described by the idea of gluon saturation within the Color Glass Condensate framework using the $k_T$-factorization. This establishes an important fact that the bulk of the produced particles in heavy-ion collisions at the LHC carries signature of the initial stage of collisions.  We also show that the recent scaling property seen by the CMS collaboration between the number of tracks in p+p and p+Pb collisions may provide a strong evidence in favor of geometric-scaling phenomenon and gluon saturation, indicating that the underlying dynamics of high multiplicity events in p+p and p+Pb collisions should be similar.

\end{abstract}

\maketitle

%---------------------------------------------------------------------------
\section{Introduction}
The recent LHC measurements of particle correlations in azimuthal and pseudorapidity in proton-lead (p+Pb) collisions \cite{r1,r2,r3}, the so-called Ridge,  and its similarity to the same effect in lead-lead (Pb+Pb) collisions, have raised the question whether the same underlying dynamics  is responsible in both cases. There are two very distinct pictures which equally provide a good description of this phenomenon,  the Color Glass Condensate (CGC) approach based on the initial-state (and gluon saturation) effects in the nucleon or nuclear wavefunctions \cite{r-cgc1} (see also Refs.\,\cite{r-cgc2,r-cgc3,r-cgc5}) and  the hydrodynamical approach based on the final-state  rescattering effects \cite{h1,h2,h3}. In order to distinguish between different scenarios of particle production mechanisms in p+Pb collisions, it is important to investigate on the equal-footing,  the multiplicity dependence of particle production, correlations, event shapes in p+p, p+Pb, and Pb+Pb collisions.  Such studies have been already undertaken by the LHC experiments.

Recently the ALICE \cite{alice-av} and the CMS \cite{cms-av} collaborations reported the  measurements of the first moment $\langle p_T\rangle$ of the charged-particle transverse momentum spectrum  and its correlation with the charged-particle multiplicity $N_{ch}$ in p+p, p+Pb, and Pb+Pb collisions.  While the rise of  $\langle p_T\rangle$ with $N_{ch}$ in p+p collisions can be understood within a model with final-state color reconnection between strings produced in multiple parton interactions \cite{cc,py}, the same model cannot describe  p+Pb and Pb+Pb data. On the other hand, while the EPOS model \cite{epos} which assumed collective flow, describes p+Pb data at high-multiplicity, it over-predicts the Pb+Pb data and gives opposite trend versus $N_{ch}$. Other Monte Carlo event generators such as  DPMJET \cite{dp}, HIJING \cite{hi} and AMPT \cite{am} also fail to describe these data, for model comparison see Ref.\,\cite{alice-av}. 
Therefore,  an unified description of  $\langle p_T\rangle$ versus $N_{ch}$ from p+p, p+Pb to Pb+Pb collisions seem to be a challenge. The main aim of this letter is to address the importance of the initial-state effect, a missing important ingredient in all above-mentioned models. Here, we provide an unified description of all these data, within the CGC approach \cite{cgc1,cgc2,cgc3} based on the gluon saturation \cite{cgc1} and Glasma physics \cite{gls-0,gls}.  We also show that the scaling property between the number of tracks in p+p and p+Pb collisions, recently observed by the CMS collaboration \cite{cms-av}, indicates the geometric-scaling phenomenon \cite{gs,gs-pb} as expected in high-energy QCD.

\section{Main formulation}

In the CGC approach \cite{cgc1,cgc2,cgc3}, the gluon jet production in A+B collisions can be described by $k_T$-factorization given by \cite{KTINC},
\begin{equation} \label{M1}
\frac{d \sigma}{d y \,d^2 \qt}=\frac{2\alpha_s}{C_F}\frac{1}{p^2_T}\int d^2 \kt \phi^{G}_{A}\Lb x_1;\kt\Rb \phi^{G}_{B}\Lb x_2;\qt -\kt\Rb,
\end{equation}
where $C_F=(N_c^2-1)/2N_c$ with $N_c$ being the number of colors. We define 
$x_{1,2}=(q_T/\sqrt{s})e^{\pm y}$ where $q_T$, $y$  are the
transverse-momentum and rapidity of the produced gluon
jet, and $\sqrt{s}$ is the collision energy per nucleon. The unintegrated gluon
density $\phi^{G}_{A}(x_i;\kt)$ gives the probability to find a gluon that carries $x_i$
fraction of energy with $k_T$ transverse momentum in the projectile
A (or target B). The unintegrated gluon
density is related to the imaginary part of the forward quark anti-quark scattering amplitude on a proton (or
nucleus) target $N_{p (A)}\Lb x_i; r_T; b \Rb$ via,  
\beq \label{M2}
\phi^G_A\Lb x_i;\kt\Rb=\frac{1}{\alpha_s} \frac{C_F}{(2 \pi)^3}\int d^2 \bt\, d^2 \rt\,
e^{i \kt\cdot \rt}\nabla^2_T N^G_A\Lb x_i; \rt; \bt\Rb,
\eeq
with a notation
\beq \label{M3}
N^G_A\Lb x_i; \rt; \bt \Rb =2 N_A\Lb x_i; \rt; \bt\Rb - N^2_A\Lb x_i; \rt; \bt \Rb,
\eeq
where $r_T$ denotes the dipole transverse size and $\bt$ is the
impact parameter of the scattering. The most important ingredient of the single inclusive hadron
production cross-section in \eq{M1} which captures the saturation dynamics is the
fundamental or adjoint dipole amplitude.  The impact-parameter dependence of the amplitude is crucial here. We use the impact-parameter Color Glass Condensate (b-CGC) dipole saturation model \cite{b-cgc}.  In the b-CGC dipole model, the color dipole-proton amplitude is given by, 
\bea \label{CA5}
N_p\Lb x, r, b\Rb\,\,=\,\, \left\{\begin{array}{l}\,\,\,N_0\,\Lb \frac{r Q_s}{2}\Rb^{2\gamma_{eff}}\,\,\,\,\,\,\,\,r Q_s\,\leq\,2\,,\\ \\
1\,\,-\,\,\exp\Lb -\mathcal{A} \ln^2\Lb \mathcal{B} r Q_s\Rb\Rb\,\,\,\,\,\,\,\,\,\,\,\,\,\,\,\ rQ_s\,>\,2\,,\end{array}
\right.
\eea
with effective anomalous dimension defined as 
\beq \label{g-eff}
\gamma_{eff}=\gamma_s\,\,+\,\,\frac{1}{\kappa \lambda Y}\ln\Lb\frac{2}{r Q_s}\Rb,
\end{equation}
where  $Y=\ln(1/x)$ and $\kappa= \chi''(\gamma_s)/\chi'(\gamma_s)$, with $\chi$ being the LO BFKL
characteristic function. The parameters $\mathcal{A}$ and
$\mathcal{B}$ in \eq{CA5} are determined uniquely from the matching of the dipole amplitude and
its logarithmic derivatives at $r Q_s=2$. The dipole amplitude depends on the saturation scale $Q_s$, namely the momentum scale at which non-linear gluon recombination effects start to become as important as the gluon radiation \cite{cgc1}. In the b-CGC dipole model, the saturation scale of proton explicitly depends on the impact-parameter and is given by
\beq \label{qs-b}
Q_s\equiv Q_{s}(x,b)\,\,=\,\,\Lb \frac{x_0}{x}\Rb^{\frac{\lambda}{2}}\,\exp\left\{- \frac{b^2}{4\gamma_s B_{CGC}}\right\} \text{GeV}. 
\end{equation}
The parameters of the b-CGC dipole amplitude ($N_0$, $B_{CGC}$, $\gamma_s, x_0, \lambda$) are determined via a fit to HERA data \cite{b-cgc}. 
This model approximately incorporates all known
properties of small-x regime of QCD \cite{bcgc-b1} including the
impact-parameter dependence of the scattering amplitude \cite{bcgc-b2} and it 
describes all existing small-$x$ data at HERA both in the inclusive and exclusive diffractive processes including the recently released high precision combined HERA data \cite{b-cgc}. 
The dipole-nuclear amplitude is constructed in the standard fashion from the dipole-proton amplitude and proton saturation scale by employing the nuclear thickness function \cite{LR-pp,LR-aa}. For a review  of $k_T$-factorization phenomenology within different saturation models see Refs.\,\cite{kt-all,kt-all2}.

In order to take account of the difference between rapidity $y$ and the
measured pseudo-rapidity $\eta$, we employ the Jacobian transformation $h(p_T, \eta, m_{jet}^2)$
between $y$ and $\eta$ where the parameter $m_{jet}$ is mini-jet mass which mimics the
pre-hadronization effect\footnote{By introducing mini-jet mass, we also regularize the $k_T$-factorization.} \cite{LR-pp}. For the value of strong-coupling $\alpha_s$ in Eqs.\,(\ref{M1},\ref{M2}) we employ the running coupling prescription used in Ref.~\cite{LR-pp,LR-aa,LR-energy}. The average transverse momentum of the mini-jet is calculated from \eq{M1}, 
\beq   \label{av}
\langle q^{\mbox{mini-jet}}_T\rangle\,=\,\int d \eta \int d^2 \qt\,h(q_T, \eta, m_{jet}^2) |q_T|\,\frac{d \sigma}{dy \,d^2 \qt}\Big{/} \int d\eta \int d^2 \qt\,h(q_T, \eta, m_{jet}^2) \frac{d \sigma}{dy \,d^2 \qt}.
\eeq
The average transverse momentum of the mini-jet can be directly related to the saturation scale, see below.

The first question one has to address here is whether via the single-inclusive gluon production, one can generate high multiplicity $N_{ch}>>\langle N_{ch} \rangle$ in p+p collisions.  One can show that with centrality cut based on the {\it single-inclusive} $k_T$-factorization supplemented with the proton-dipole amplitude, one can maximally generate upto $N_{ch}\approx 1\div 2\langle N_{ch} \rangle $ for very central p+p collisions at $\sqrt{s}=7 $ TeV and $\eta=0$. Although the main underlying mechanism of high-multiplicity event production in p+p and p+A collisions is still unknown, multi-gluon production and fluctuations in geometry and color charge distributions should play significant role. In the CGC framework, the yield of multiple gluons production can be obtained from Glasma color flux tubes decay \cite{gls} which leads to the negative binomial distribution for the multiplicity, in accordance with experimental observation at RHIC and the LHC \cite{raj}.  Using the multiple gluons production yield from Glasma flux tubes decay \cite{gls}, one can immediately show that in leading log approximation, deep in the gluon saturation regime, the average transverse momentum of a produced gluon in multiple-gluon production is equal to computing the same quantity in the single-inclusive production, 
\begin{equation}\label{gl-2}
\left\langle \left\langle \qt \right\rangle\right\rangle=\left\langle \qt \right\rangle,
\end{equation}
where we used the notation $\left\langle \left\langle \dots \right\rangle\right\rangle$ to indicate the averaging in multiple-gluon yield while the notation $ \left\langle \dots \right\rangle$ is used for a averaging over a single-inclusive yield. In the Glasma picture, the transverse area $S_\perp$ is filled with $Q_s^2S_\perp$ independent flux tubes of size $1/Q_s^2$ and each of these tubes emits gluons with  the typical momentum of order the saturation scale $Q_s$. It may then seem natural to expect that \eq{gl-2} to be held deep inside the saturation region. Therefore, at high-multiplicity events at the LHC, assuming that we are in the saturation region,  the average transverse momentum of a gluon-jet  can be directly obtained via the single-inclusive $k_T$-factorization defined in \eq{av}, even though, the multiple-gluon production is the main source behind the production of such rare events. This may also be considered here as our working hypothesis. 

In the CGC picture, the gluon saturation scale is proportional to the parton density, and  since the parton density is proportional to the particle multiplicity, then the saturation scale appeared in \eq{gl-2} should depend on the multiplicity, its own first cause. Let us define two saturation scales: one for the projectile $Q_s^A$ and another for the target $Q_s^B$, we also introduce two auxiliary variables $Q_{s, min}=min.\{Q_s^A,Q_s^B\}$, and $Q_{s, max}=max.\{Q_s^A,Q_s^B\}$. 
Using the $k_T$-factorization \eq{M1}, one can show that deep in the saturation region where $q_T<Q_{s,min}$ we approximately have,
\begin{equation}\label{kt-n}
\frac{dN}{dy}\propto S_{\perp} Q_{s, min}^2,
\end{equation}
while in the kinematic region where only one of target or projectile is in the saturation region $Q_{s,max}>q_T>Q_{s,min}$, within the same approximation we have \cite{cgc-app}, 
\begin{eqnarray} \label{kt-n2}
\frac{dN}{dy}\propto S_{\perp} Q_{s, min}^2\ln^2\frac{Q_{s, max}^2}{Q_{s, min}^2},\nonumber\\
\langle q_T\rangle \propto  Q_{s,max}\frac{\zeta-1-\ln\zeta}{\ln^2\zeta},\
\end{eqnarray}
where $\zeta= Q_{s,min}/ Q_{s,max}$.  Note that the above relations are by construction built in the KLN model \cite{kln}. Here, in accordance with the underlying assumption in \eq{gl-2}, we consistently assume that we are deep inside of saturation regime and take the relation given in  \eq{kt-n} to relate the saturation scale to charged-particle multiplicity $N_{ch}$ by fixing the prefactor in \eq{kt-n} to the average minimum-bias charged-particle multiplicity $\langle N_{ch} \rangle$. Therefore, we replace
\begin{equation}\label{s-nc}
Q^2_{s, min}\to \,\frac{N_{ch}}{\langle N_{ch} \rangle}\,Q^2_{s,min}, 
\end{equation}
 where the integral in $b$ is now in minimum-bias\footnote{We checked that for $N_{ch}<\langle N_{ch} \rangle$ numerically our prescription is equivalent to introducing centrality dependence by imposing cut in impact-parameter.  The advantage of  introducing a $N_{ch}$-dependent saturation scale is that it is free  from possible uncertainties associated with centrality-cut definition in p+p and p+A collisions. }.  We recall that large multiplicity events  with $N_{ch}>>\langle N_{ch} \rangle$  can be described by the multiple gluon production yield obtained from Glasma flux decay yield, see the discussion after \eq{av}. On the other hand, the average transverse momentum of the produced single inclusive gluon in a high-multiplicity event can be still obtained via the $k_T$-factorization \eq{M1} due to the equality given in \eq{gl-2}.  Note that the saturation scale given in \eq{s-nc} by construction reproduces the charged-particle multiplicity $N_{ch}$ via the $k_T$-factorization \eq{kt-n} upto some logarithmic corrections (consistent with the same approximation made in \eq{gl-2}). In this sense, deep inside of the saturation regime, one can approximately mimic the effects of Glasma flux decay (and fluctuations) by redefining the saturation scale and employ \eq{M1} to  compute the average transverse momentum  of the produced gluon in such rare high-multiplicity events.

It is instructive to note that from Eqs.\,(\ref{kt-n},\ref{kt-n2}), for symmetric collisions like p+p and A+A collisions at mid-rapidity with $Q_{s, min}=Q_{s, max}$, we  approximately have
\begin{equation} \label{kt-nq}
\langle q_T\rangle \propto (\frac{1}{S_{\perp}}\frac{\ud N}{\ud y })^{1/2},
\end{equation}
 while in asymmetric cases like p+A collisions where one of the sources is significantly weaker than the other one with $|\ln \zeta|>>1-\zeta$, the multiplicity dependence of $\langle q_T\rangle$ becomes weaker than the former cases and the simple relation in \eq{kt-nq} is not more reliable \cite{cgc-app}. 
This is consistent with the fact that based on only dimensionality consideration, the relation in \eq{kt-nq} is approximately expected to be held in the saturation region.  However, it is important to note that at a fixed multiplicity the interaction area in p+p, p+A and A+A collisions are very different and in practice there are large logarithmic corrections to \eq{kt-nq}. Our main aim here is to calculate $\langle q_T\rangle$  via \eq{av} without resorting to any  approximation.

\begin{figure}[t]                            
                           \includegraphics[width=8 cm] {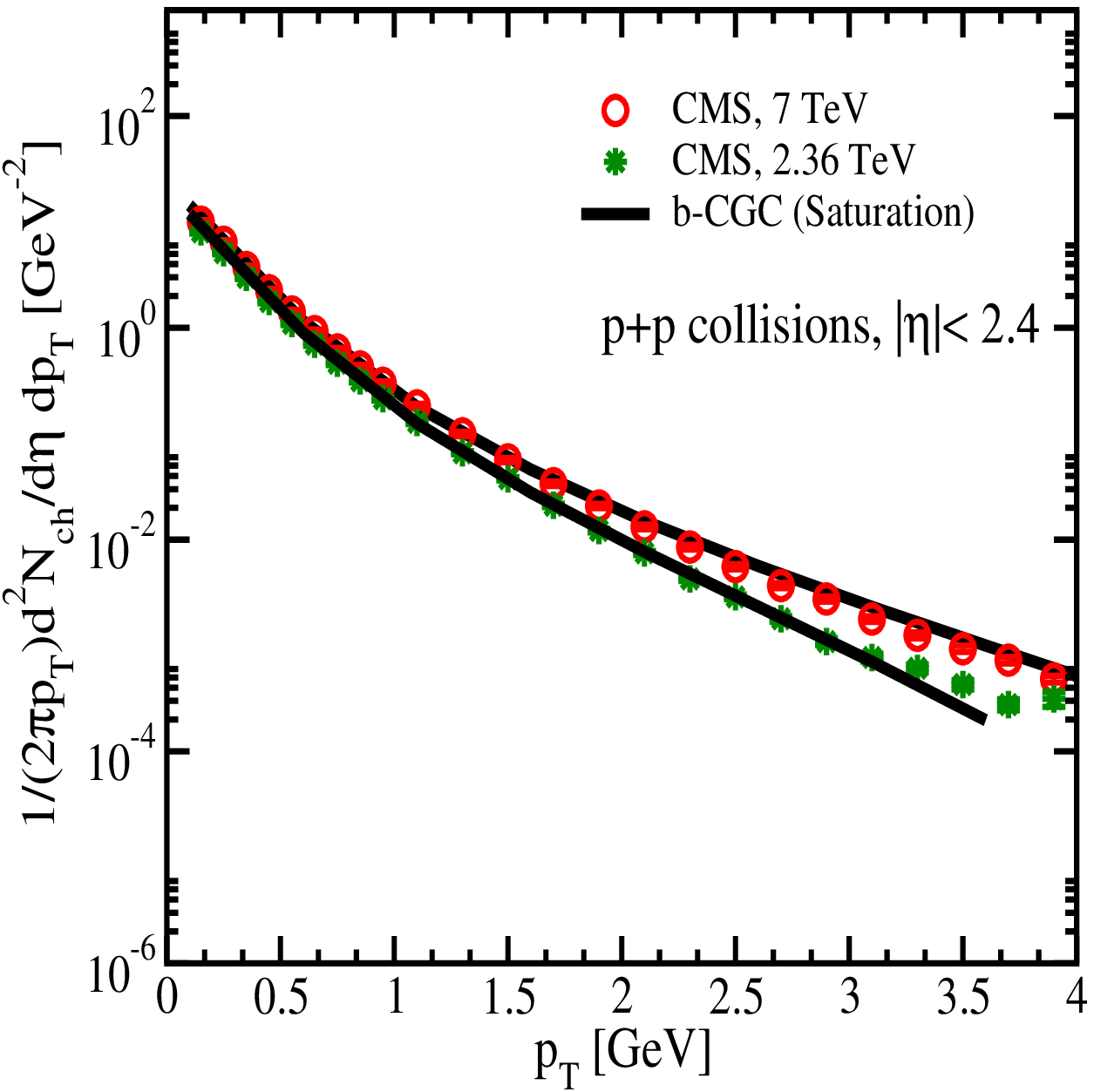}
                                  \includegraphics[width=8 cm]
                                  {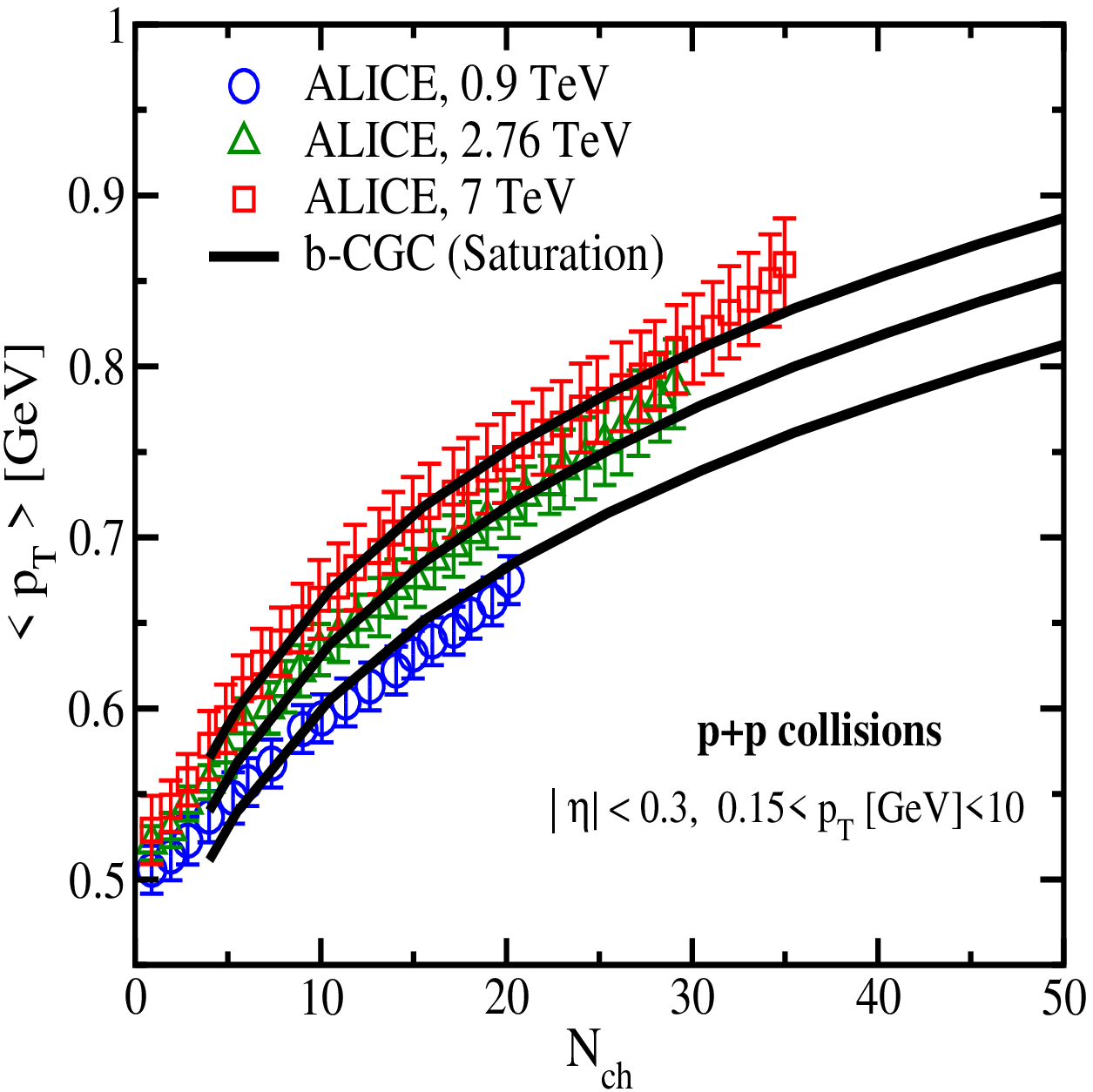} 
\caption{Left: The differential yield of charged hadrons in p+p collisions  at $\sqrt{s}=2.36 $ and $7$ TeV for $|\eta|<2.4$. The experimental data are from the CMS collaboration \cite{cms-pp}.  Right: Average transverse momentum $\langle p_T\rangle$ in the range $0.15<p_T<10$  GeV as a function of charged-particle multiplicity $N_{ch}$ in p+p collisions at $\sqrt{s}=0.9,2.76 $ and $7$ TeV for $|\eta|<0.3$. The experimental data are from the ALICE collaboration \cite{alice-av}. The black curves in both panel are results from the Color Glass Condensate based on the $k_T$-factorization and the b-CGC dipole saturation model. }
\label{fig1}
\end{figure}

Finally, we should also specify our hadronization scheme. It is generally an open problem how to incorporate the fragmentation processes into the CGC/saturation formalism. 
In order to clearly disentangle the initial from final-state effect, we employ here a simple scheme for  the final-state hadronization, 
based on the so-called Local Parton-Hadron Duality (LHPD) principle \cite{mlla}. Namely we assume that the hadronization is a soft (or semi-soft) process
and cannot change the direction of the emitted radiation. Furthermore, we assume that the hadronization occurs in vacuum and it is the same for p+p, p+Pb, and Pb+Pb collisions.  Note that the main contribution of the $k_T$-factorization for
the multiplicity distribution comes from small transverse momentum $p_T < 1$ GeV. This is in accordance with the experimental observation that the average transverse momentum of the produced charged-particle is rather small $\langle p_T\rangle<1$ GeV for a wide range of kinematics and system size, see Figs.\,\ref{fig1},\ref{fig2}. 
Note that the fragmentation functions based on fixed order approximation to DGLAP evolution is dubious at low transverse momentum\footnote{For the same reason both, the AKK \cite{akk} and the KKP \cite{kkp} fragmentation functions, are only given for $Q>1$ GeV.} and we do not employ them here since we are only interested in low $p_T$ region. 
In the framework of the LHPD, the $p_T$ spectrum of the produced hadron upto a normalization, is obtained from \eq{M1} by replacing $q_{ T}=p_T/\langle z  \rangle$ where $p_T$ is the transverse momentum of the produced hadron and the parameter $\langle z  \rangle$ is the average fraction of energy of the mini-jet carried by the hadron. The average transverse momentum of the produced hadrons can be then obtained by
\begin{equation} \label{av-z}
\langle p_{T} \rangle = \left( \langle \langle z  \rangle\,q^{\mbox{mini-jet}}_T\rangle^2+\langle p^{\mbox{intrinsic}}_T \rangle^2\right)^{1/2},
\end{equation}
where $\langle p^{\mbox{intrinsic}}_T\rangle$ is the average intrinsic
transverse momentum of the hadron in the mini-jet and $\langle\,q^{\mbox{mini-jet}}_T\rangle$ is computed via \eq{av}. 

\section{Numerical results and discussions}
In our formalism, we have only four unknown parameters, the overall normalization factor in \eq{M1}, the mini-jet mass $m_{jet}$, the average fragmentation parameter $ \langle z  \rangle$ and  the average intrinsic transverse momentum of the hadron $\langle p^{\mbox{intrinsic}}_T\rangle$. Note that for spectra of hadron, the free parameter $\langle p^{\mbox{intrinsic}}_T\rangle$ does not enter into the calculation while for the average transverse momentum the normalization will drop out. Therefore, for all observables considered here we practically have only 3 free parameters. All these unknown parameters are fixed via a fit to minimum-bias  p+p data at low-energy \cite{LR-pp,me-kt}.  Therefore, our results shown here at higher energies and different system sizes from p+p to p+A and A+A collisions should be considered as a free-parameter calculation. 
The over-all normalization, and  the mini-jet mass 
are fixed via a fit to the experimental data on the charged-particle multiplicity at mid-rapidity.
Unfortunately, we do not know how mini-jet mass changes with the system size and kinematics. Here, we focus on kinematics around mid-rapidity and assume that the mini-jet mass is independent of kinematics \cite{me-kt}. The parameters $\langle z  \rangle$  and $\langle p^{\mbox{intrinsic}}_T\rangle$  are fixed via a fit to a spectra  and average transverse momentum of charged-particle in minimum-bias p+p collisions at low energies (not shown here). We therefore obtain  $\langle z  \rangle\approx 0.5\div 0.6$, and  $\langle p^{\mbox{intrinsic}}_T\rangle_{\text{proton}}\approx 0.2\div 0.45$ GeV  for proton and nuclear target\footnote{In the CGC picture in contrast to the standard collinearly pQCD approach,  the target is described by a classical gluon field representing a multi-gluon state with collectively intrinsic transverse momentum proportional to the saturation scale $Q_s$ rather than an individual gluon with a well defined energy fraction and zero transverse momentum. Therefore, we have  $\langle p^{\mbox{intrinsic}}_T\rangle_{\text{proton}}\approx \mu Q_s\approx 0.2$ GeV where $\mu$ is a dimensionless parameter coming from soft physics. On the other hand, the saturation scale of heavy nuclei is bigger than proton with a factor roughly about $A^{1/6}$ (where $A$ is atomic number). Therefore, we expect $\langle p^{\mbox{intrinsic}}_T\rangle$ for the case of lead target to be $A^{1/6}$ bigger than the same quantity for the proton, namely $\langle p^{\mbox{intrinsic}}_T\rangle_{\text{nucleus}}\approx A^{1/6} \langle p^{\mbox{intrinsic}}_T\rangle_{\text{proton}}\approx 0.45$ GeV. In this way, one can reduce the uncertainties associated with the value of $\langle p^{\mbox{intrinsic}}_T\rangle$. }.

\begin{figure}[t]                            
                              \includegraphics[width=8 cm] {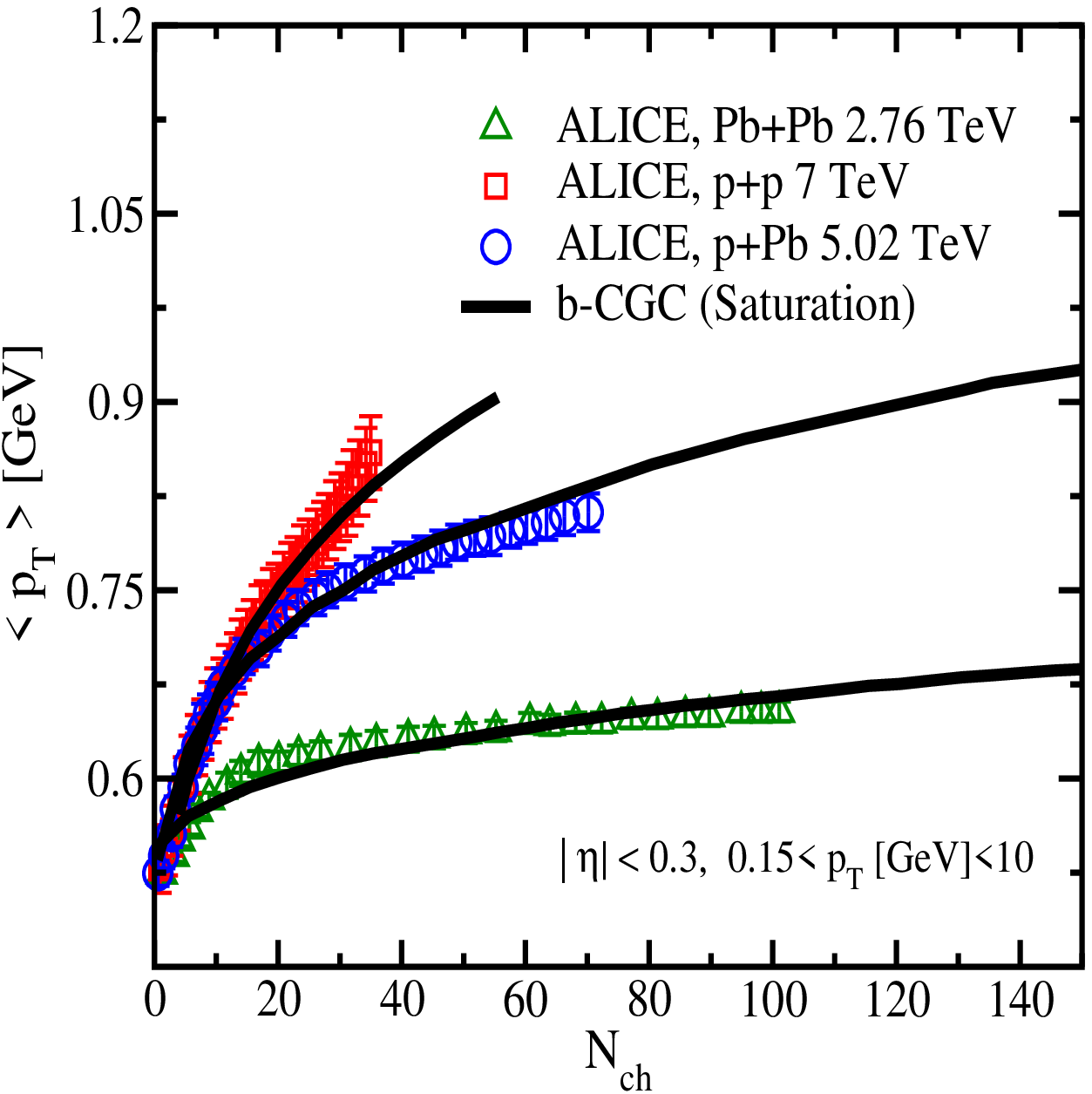}
                                  \includegraphics[width=8 cm]
                                  {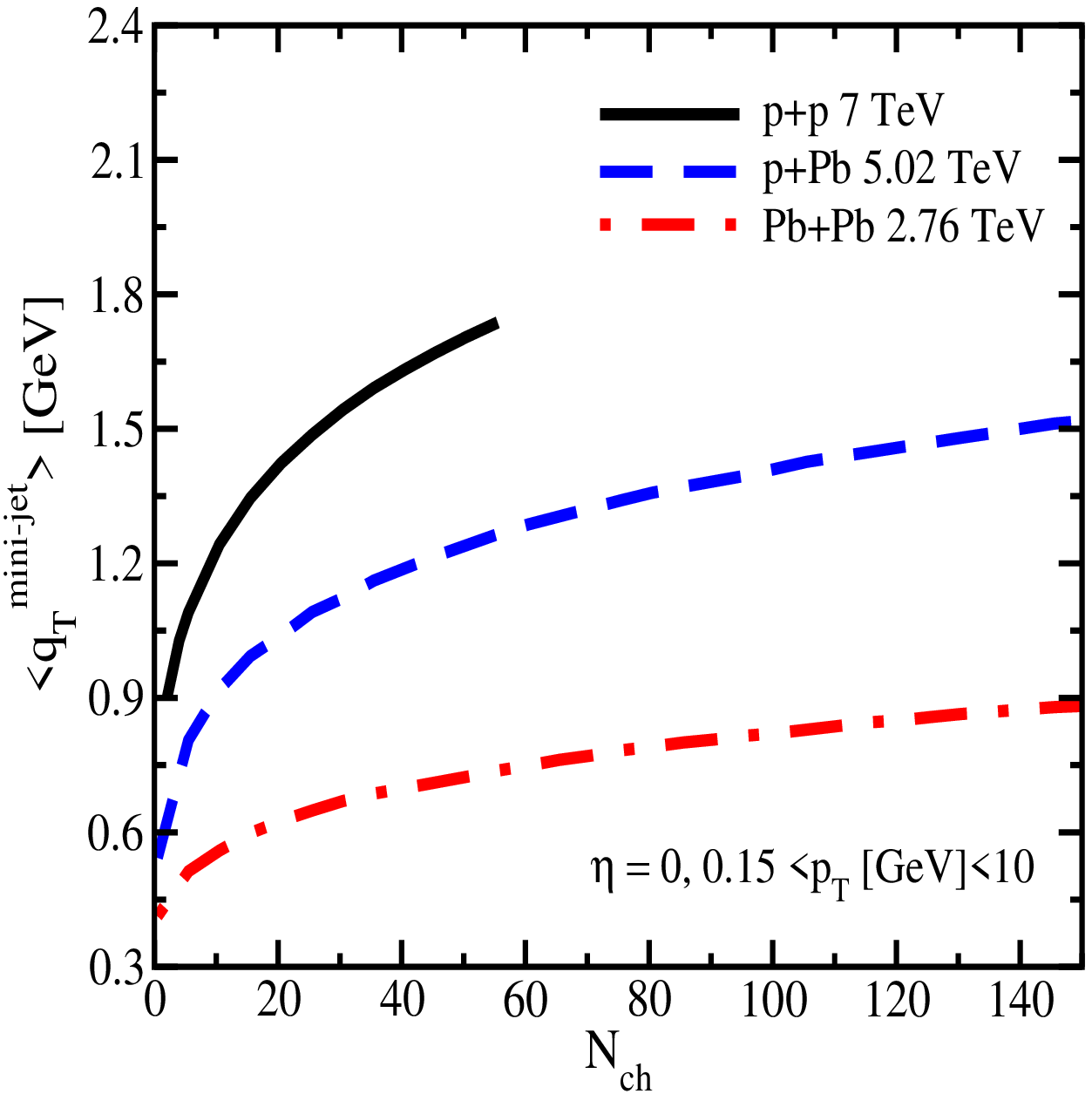}                               
\caption{Left: Average transverse momentum  $\langle p_T\rangle$ of charged particles in the range $0.15<p_T<10$ GeV as a function of charged-particle multiplicity $N_{ch}$ in p+p , p+Pb, Pb+Pb collisions at $\sqrt{s}=7, 5.02 $ and $2.76$ TeV  respectively for $|\eta|<0.3$.  
The experimental data are from the ALICE collaboration \cite{alice-av}. Right: Average transverse momentum  $\langle q^{\mbox{mini-jet}}_T\rangle$ of mini-jet as a function of  $N_{ch}$, with the transverse momentum of the produced hadron in the range $0.15<p_T<10$ GeV  in p+p , p+Pb and Pb+Pb collisions at $\sqrt{s}=7, 5.02 $ and $2.76$ TeV  respectively for $|\eta|<0.3$.   }
\label{fig2}
\end{figure}

\begin{figure}[h!]                            
                              \includegraphics[width=8 cm] {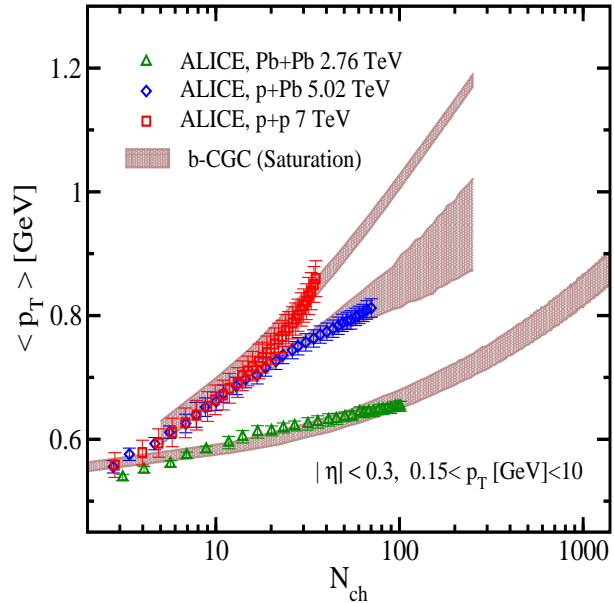}                 
\caption{Average transverse momentum  $\langle p_T\rangle$ of charged particles in the range $0.15<p_T<10$ GeV as a function of charged-particle multiplicity $N_{ch}$ in p+p , p+Pb, Pb+Pb collisions at $\sqrt{s}=7, 5.02 $ and $2.76$ TeV  respectively for $|\eta|<0.3$.  The band labeled b-CGC is consistent with the black curves shown in \fig{fig2} and includes the theoretical uncertainties, see the text for the details.  
The experimental data are from the ALICE collaboration \cite{alice-av}.   }
\label{fig2-new}
\end{figure}

In \fig{fig1} left panel, we show our results for the spectra of charged-hadron production in minimum bias p+p collisions at  $\sqrt{s}=2.36$ and 7 TeV. 
The fact that the shape of spectra of produced hadron at low $p_T$ resembles the spectra of mini-jet spectra upto a normalization provides the first hint toward importance of the initial-state effect. Note that the curve at 7 TeV in \fig{fig1} (left panel) was predicted in Ref.\,\cite{LR-pp}, see also Ref.\,\cite{pp-compare}. 

In order to make prediction for the multiplicity dependence of the average charged-hadron transverse momentum, we should also know the value of  the average multiplicity in the minimum bias event selection $\langle N_{ch} \rangle$ at various energies for different interactions. We impose the same kinematic constrains as the ones employed in the ALICE measurements \cite{alice-av}, namely we restrict the integrals in \eq{av} to $0.15<p_T<10$ GeV and $|\eta|<0.3$. In the case of p+Pb collisions, similar to the experiments at the LHC, we also take into account the rapidity shift $\Delta y=-0.465$.  As a first internal check, we reproduced  the experimental values of $\langle N_{ch} \rangle$ reported by the ALICE collaboration given in table 2 of Ref.\,\cite{alice-av}. In \fig{fig1} right panel, we compare our results with the ALICE recent data on the average transverse momentum of charged particles as a function of multiplicity in p+p collisions at various energies. Note that ALICE data \cite{alice-av} is consistent with the corresponding CMS \cite{cms-av} and ATLAS \cite{atlas-av}  measurements. The increase of $\langle p_T\rangle$ with energy and multiplicity is naturally expected in the gluon saturation picture. This is simply because $\langle p_T\rangle$ increases with the saturation scale, the only dynamical scale available in the system, and the saturation scale grows with energy and density, see \eq{kt-nq}.

 In \fig{fig2} left panel, we show average transverse momentum  of charged-particle as a function of charged-particle multiplicity $N_{ch}$ in p+p, p+Pb and Pb+Pb collisions at $\sqrt{s}=7, 5.02 $ and $2.76$ TeV  respectively, in the range $0.15<p_T<10$ GeV and $|\eta|<0.3$. In \fig{fig2} right panel similar to left panel, we show the corresponding average transverse momentum of the mini-jet  in p+p, p+Pb and Pb+Pb collisions. Comparing both panels in \fig{fig2}, it is seen that remarkably the trend of the average transverse momentum of charged particles as a function of $N_{ch}$ at various system size and energies resembles the same quantity for the mini-jet gluon.  The average transverse momentum of charged particles  $\langle p_T\rangle$  is smaller in Pb+Pb than p+Pb and p+p collisions at high-$N_{ch}$.  The main reason is that the effective interaction area is different in Pb+Pb compared to p+p and p+Pb collisions. Note that events with $N_{ch}<\langle N_{ch} \rangle$, is more prepherial and less dense compared to the minimum bias collisions. We recall that the average charged-particle multiplicity measured by the ALICE collaboration \cite{alice-av} is $\langle N_{ch} \rangle\approx 259.9, 11.9$ and $4.42$ in Pb+Pb, p+Pb and p+p collisions respectively at the kinematics considered in \fig{fig2}.  Therefore, a event with a given value of $N_{ch}$ in \fig{fig2}  corresponds to different density in p+p, p+Pb and Pb+Pb collisions. In other words, at moderate $N_{ch}$ in range of $N_{ch}<150$ shown in \fig{fig2} we are in dilute region in Pb+Pb collisions given that $\langle N_{ch} \rangle\approx 259.9$ while the same multiplicity event selection corresponds to a very rare high-density event in p+p collisions.  Note that the ALICE data in Pb+Pb collisions shown in Figs.\,\ref{fig2},\ref{fig2-new}  correspond to peripheral collisions with  $\langle N_{part}\rangle <52$  \cite{alice-c, atlas-c,cms-c}. We recall that at $50-60\%$ centrality in $|\eta|<0.5$ we have $dN_{ch}/d\eta =149\pm 6$ and $\langle N_{part}\rangle=52.8\pm 2.0$ \cite{alice-c}.

The rise and flatness of $\langle p_T\rangle$ in different collisions with different system sizes can be also simply understood from \eq{kt-nq}. First, we recall that in the saturation region, the interaction area is subject to rapid rise with multiplicity and then it becomes independent of multiplicity and flattens at a critical $N_{cri}$ which is typically larger than  $\langle N_{ch} \rangle$ \cite{ip-gls1}. Now, given the fact that  we have 
\begin{eqnarray}
&& \langle N_{ch} \rangle^{Pb+Pb}> \langle N_{ch} \rangle^{p+Pb} > \langle N_{ch} \rangle^{p+p}, \\
&& N_{cri}^{Pb+Pb}>N_{cri}^{p+Pb}>N_{cri}^{p+p}, \
\end{eqnarray}
at low $N_{ch}<\langle N_{ch} \rangle$, increase in multiplicity and  the interaction area, approximately cancel out each others in \eq{kt-nq}  leading  to the flatness of the average transverse momentum. In the case of p+p collisions with  $N_{ch}>>\langle N_{ch} \rangle$, shown in \fig{fig2}, the interaction area is roughly constant and then the average transverse momentum increases with multiplicity as expected from \eq{kt-nq}, in accordance with experimental data. Similar to \fig{fig2}, in  \fig{fig2-new} we show the average transverse momentum of charged particles as a function of charged-particle multiplicity $N_{ch}$ in a wider range of $N_{ch}$, upto $N_{ch}=1400$ for Pb+Pb collision at the LHC. In order to clearly see the slow rise of $\langle p_T\rangle$ with $N_{ch}$, we plotted the results in logarithmic scale.  The bands labeled b-CGC is consistent with the curves shown in \fig{fig2} and include the theoretical uncertainties associated to fixing the mini-jet mass $m_{jet}$ in \eq{av}, and the parameters $\langle z\rangle$ and $\langle p^{\mbox{intrinsic}}_T \rangle$ in \eq{av-z}.

\begin{figure}[t]                             
 \includegraphics[width=8 cm] {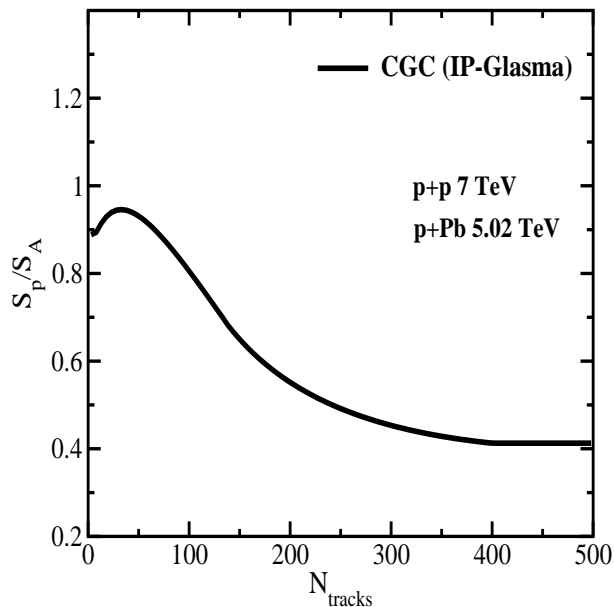}                    
\caption{ The ratio of interaction areas in p+p $\sqrt{s}=7$ TeV and p+Pb $\sqrt{s}=5.02$ TeV  as a function of tracks charged-particle multiplicity for $| \eta |<2.4$  obtained from the IP-Glasma model \cite{ip-gls1}. }
\label{fig-s}
\end{figure}

\begin{figure}[t]              
  \includegraphics[width=8 cm] {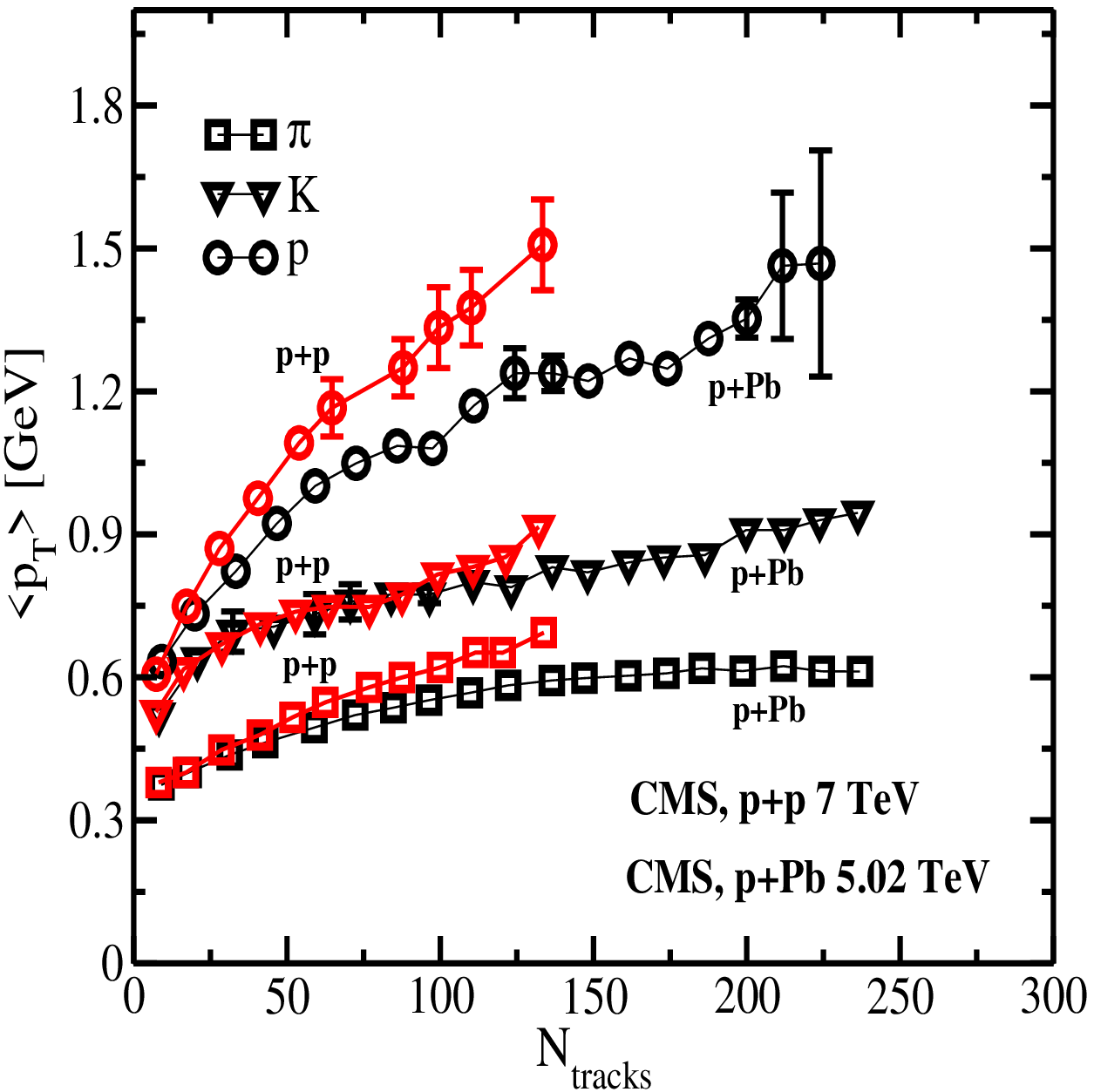}     
                \includegraphics[width=8 cm] {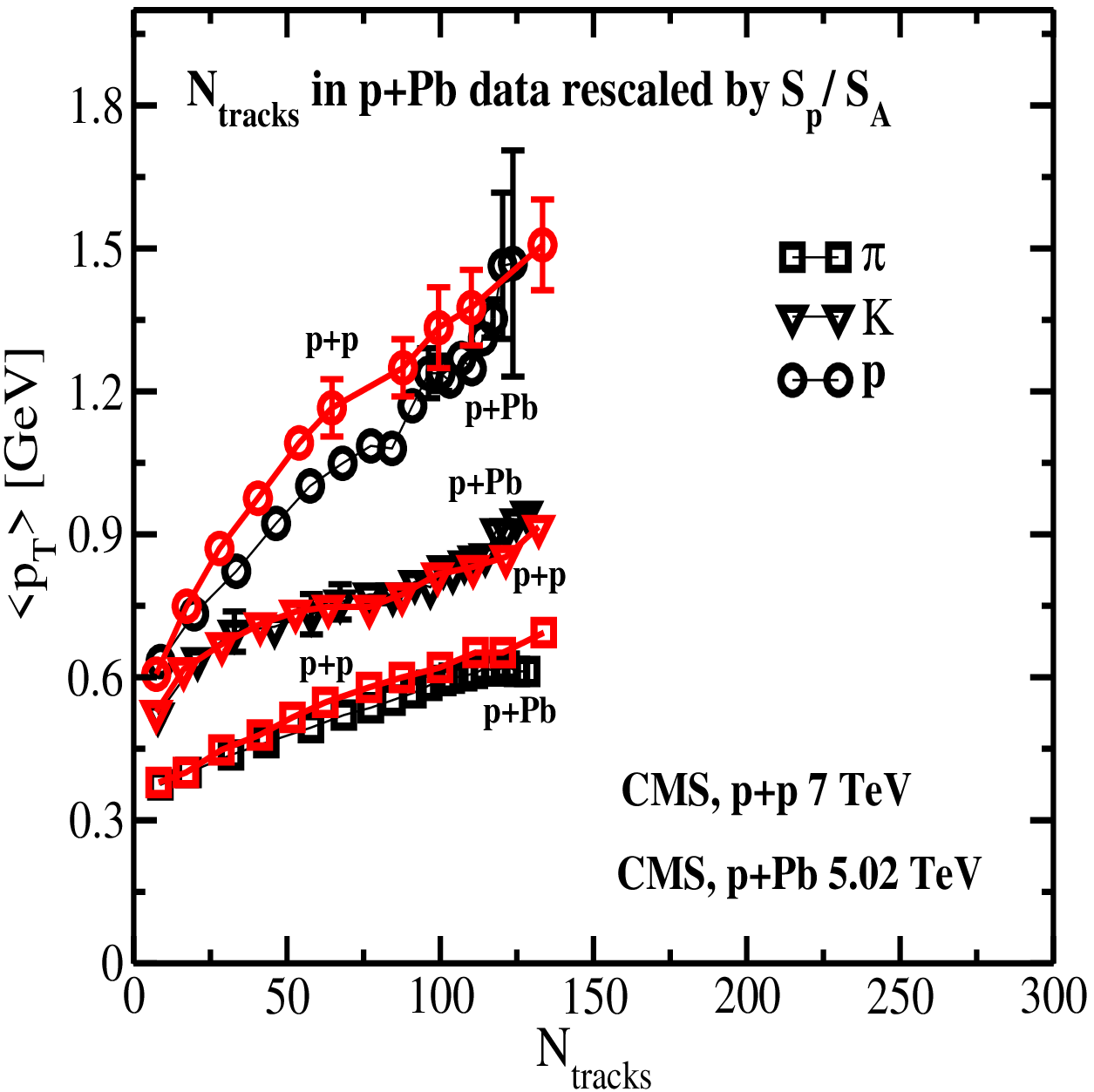}  
\includegraphics[width=8 cm] {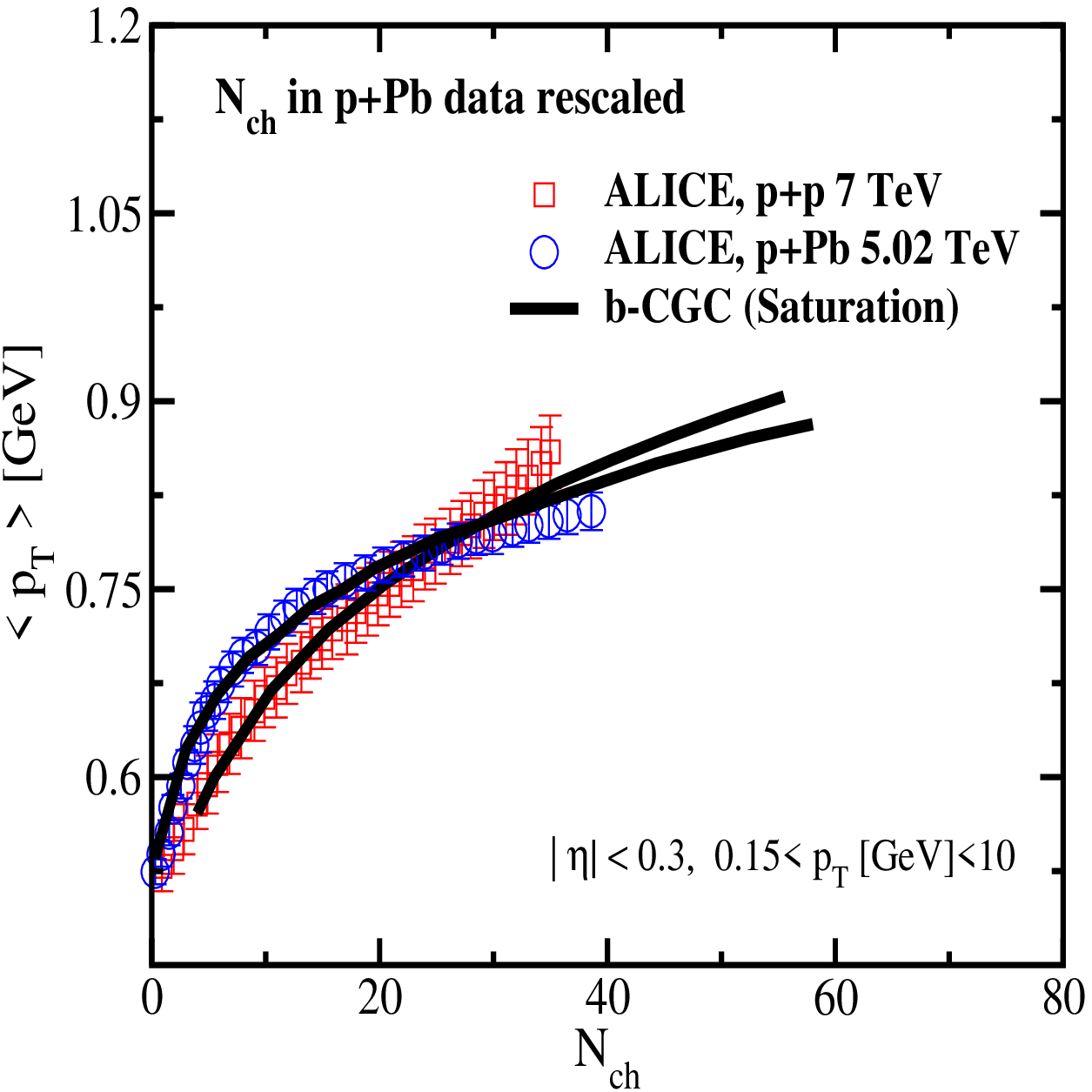}                                   
\caption{Upper left: average transverse momentum of identified charged hadrons (pions, kaons, protons) in the range $| y |<1$, for all particle types, as a function of the corrected track multiplicity for $| \eta|<2.4$, for pp collisions at $\sqrt{s}=7$ TeV, and for p-Pb at $\sqrt{s}=5.02$ TeV measured by the CMS collaboration. Upper right: 
The same as left with $N_{tracks}$ values scaled out via \eq{sg} with the ratio of interaction areas shown in \fig{fig-s}, see the text for the details. Lower: 
 $\langle p_T\rangle$ as a function of charged-particle multiplicity $N_{ch}$ in p+p  and  p+Pb collisions at the LHC with values of $N_{ch}$ in p+Pb collisions rescaled.  
The experimental data are from Refs.\,\cite{alice-av,cms-av}.}
\label{fig3}
\end{figure}

The CMS collaboration \cite{cms-av} has recently observed a very intruging scaling behaviour  in average transverse momentum of various identified hadrons as a function of the true track multiplicity $N_{tracks}$  in p+p and p+Pb collisions for $|\eta|<2.4$. Namely it was found that the p+Pb curves (in left panel \fig{fig3}) is approximately similar to p+p curves by taking the p+p values and multiplying their $N_{tracks}$ coordinate by a factor of 1.8, for all particle types, see \fig{fig3}. In other words, p+Pb collisions with a given $N_{tracks}$ is similar to a p+p collisions with $0.55\times N_{track}$. It is important to notice that this scaling is the same for all different identified hadrons indicating that perhaps its origin may be traced back to the initial stage of collisions before hadronization. The observed scaling in number of tracks is in fact expected in the CGC approach. We recall that in the CGC framework, two multiplicity events with the same saturation scale should lead to the same physics. Now, using \eq{kt-n}, one can immeaditely relate track multiplicity in p+p and p+Pb collisions with a similar saturation scale: 
\begin{equation} \label{sg}
 N_{tracks}^{p+p}  \propto \,S_p Q^2_{s,min}\Lb x \Rb\,\,\,\, \text{and}\,\,\,\, N_{tracks}^{p+Pb}  \propto \,S_A Q^2_{s,min}\Lb x \Rb \longrightarrow N_{tracks}^{p+p}\approx  K \frac{S_{p}}{ S_A}\, N_{tracks}^{p+Pb}, 
\end{equation}  
where $N_{tracks}^{p+p}$ and $N_{tracks}^{p+Pb}$ denote track multiplicity in p+p and p+Pb collisions respectively. $S_p$ and $S_A$ are the effective transverse interaction areas for a given centrality or multiplicity in p+p and p+A collisions, respectively.  The pre-factor $K$ takes into account the effect of different center-of-mass energy  in p+p and p+Pb collisions at the LHC. The saturation scale of proton is  $Q_s^2\propto (x_0/x)^\lambda \propto s^{\lambda/2}$, where the parameter $\lambda\sim 0.22$ was extracted from the LHC data in p+p collisions \cite{LR-energy}. Therefore we have
$K\sim \left( \sqrt{s}=7\,\text{TeV}/\sqrt{s}=5.02\,\text{TeV}\right)^{\lambda}=1.076$. For calculating the interaction area as a function of tracks (or multiplicity), we use the recent results of the IP-Glasma initial-state model of hadrons and nuclei \cite{ip-gls1}.
The IP-Glasma model \cite{ip-gls1,ip-gls2}, is a CGC-type approach to particle production at early stage of collisions based on the classical Yang-Mills description of initial Glasma fields which properly incorporates the impact parameter dependence via the IP-Sat dipole model for proton \cite{ip-sat}. In the IP-Glasma \cite{ip-gls1}, one can from the first principle compute the radius of interaction in term of gluon multiplicity. In order to relate the gluon multiplicity to the number of corrected tracks, following Ref.\,\cite{gs-pb} we employ 
\begin{equation}
\frac{dN_g}{dy}\approx \frac{3}{2}\frac{1}{\Delta\eta}N_{tracks}, \label{track}
\end{equation} 
where for the CMS kinematics of interest with $|\eta|<2.4$, we take  $\Delta\eta=4.8$.  In \fig{fig-s}, we show the ratio of interaction areas in p+p $\sqrt{s}=7$ TeV and p+Pb $\sqrt{s}=5.02$ TeV collisions as a function of number of tracks.  Note that the scaling-factor in \eq{sg} (shown in \fig{fig-s} up to the pre-factor $K$) changes slowly with $N_{tracks}$ with a median about $0.55$,  and at very large $N_{tracks}$ it is constant since at very high-multiplicity the interaction areas in both p+p and p+A collisions do not change further. 
In \fig{fig3} left panel, we show first average transverse momentum of identified charged hadrons (pions, kaons, protons) in the range $| y |<1$, for all particle types, as a function of the track multiplicity for $|\eta|<2.4$ in p+p and p+Pb collisions. 
Using $S_p/S_A$ shown in \fig{fig-s},  we can relate the CMS number of tracks $N_{tracks}$ (and ALICE charged-particle multiplicity $N_{ch}$) in p+p and p+A collisions via \eq{sg}.  In \fig{fig3},  we rescaled track multiplicity in p+Pb collisions via  \eq{sg}, and compare the average transverse momentum of various identified charged hadrons in p+p and p+Pb collisions. In \fig{fig3} (lower panel), we demonstrate that the ALICE data shown in \fig{fig2}, follows the same scaling property seen in the CMS data, within error bars. In \fig{fig3} (lower panel), we also compare our results obtained from the $k_T$-factorization having rescaled the results in p+Pb collisions. It is interesting to see that the rescaled full calculation results agree very well with the rescaled data. This indicates that the small discrepancy seen in \fig{fig3} between p+p and rescaled p+Pb data, is mainly due to approximation employed in \eq{sg}. Nevertheless, it is remarkable that simple geometric-scaling relation in \eq{sg}, without introducing any new parameters or ingredients, is in accordance with data for identified charged hadrons and charged particles. Note that the key ingredient in \eq{sg} is the saturation scale $Q_{s,min}$ (the proton saturation scale) which relates charged-particle multiplicity in p+p and p+Pb collisions. The interaction areas and the over-all normalization factor in \eq{sg} are then computed within the saturation model.

There are, however, a number of caveats which need further study before taking the number predicted here for the scaling-factor at face value. First, note that in the above, for simplicity we have ignored the impact-parameter $b$ dependence of the saturation scale and possible correlation between $x$ and $b$.  Therefore, the relation given in \eq{sg} is less reliable for prepherial collisions at low  $N_{tracks}$ where a proper treatment of impact-parameter dependence of the collisions is indispensable. Moreover, in Eqs.\,(\ref{sg},\ref{track}),  we approximately related the multiplicity to number of track with a constant factor and assumed that this relation to be the same for both p+p and p+A collisions. One should also bear in mind that the ratio of interaction areas computed here in the IP-Glasma model \cite{ip-gls1,ip-gls2}, is based on the IP-Sat model \cite{ip-sat}. It will be of great interest to compute the ratio of interaction areas in the Glasma model based on the b-CGC saturation model \cite{b-cgc}. 

 Some words of caution are in order here. Strictly speaking our formalism is less reliable for p+p (and A+A) collisions at around
mid-rapidities. This is due to the fact that the $k_T$-factorization is valid for asymmetric dilute-dense collisions  \cite{KTINC,gls}
like p+A (with a very large nuclei) or p+p (and A+A) collisions at forward rapidities.

 Note that the existence of the geometric-scaling in  p+Pb collisions was also discussed in Ref.\,\cite{gs-pb}. The ALICE collaboration later confirmed this with rather large error bars \cite{alice-av}. However, the above-mentioned caveats are also presents in previous studies. Therefore further investigations on this line are needed in order to pin down the true nature of the observed scaling behaviour. It would be of great interest to see if final-state type approaches like hydrodynamic can also explain this scaling phenomenon. In principle, the scaling property in number of tracks between p+p and p+Pb collisions given in \eq{sg}, should be also correct for multiplicity dependence of  other observables in p+p and p+Pb collisions at high-energy (and low $p_T$) in the saturation region. This can be verified at the LHC.

To summerize, in this letter within the CGC framework we provided an unified description of the recent ALICE (and CMS) data on $\langle p_T\rangle$ at various energies and system sizes from p+p to p+Pb and Pb+Pb collisions. In our approach neither final-state hadronization nor collective hydrodynamic-type effects are important to describe the main features of the current data. This clearly indicates that the underlying dynamics of particle production at small-x region is universal and the main behaviour of the bulk of  the produced particles in heavy-ion collisions at the LHC can be  simply described by the idea of gluon saturation.  In \fig{fig2-new}, we show our predictions for $\langle p_T\rangle$  in a wider range of $N_{ch}$. In order to discriminate among difference models and pin down the importance of final-state effects like hydrodynamic,   it is crucial to have experimental date at large $N_{ch}$. We also showed that the observed scaling property seen in the CMS and the ALICE data between the number of tracks in p+p and p+Pb collisions may provide a strong evidence of the gluon saturation and the so-called geometric-scaling, indicating that the underlying dynamics of high multiplicity events in p+p and p+Pb should be similar.

%---------------------------------------------------------------------
\begin{acknowledgments}
The author would  like to thank Prithwish Tribedy and Raju Venugopalan for helpful discussions. The author would also like to thank Anton Andronic and Julia Velkovska for useful communication. 
This work is supported in part by Fondecyt grants 1110781.

\end{acknowledgments}

%---------------------------------------------------------------------

\end{document}